\documentclass[aps,prc,twocolumn,amsmath,amssymb,superscriptaddress,showpacs,showkeys]{revtex4}
\usepackage{dcolumn}
\usepackage{graphicx,color}
\usepackage{multirow}
\usepackage{textcomp}
\usepackage{physics}
\usepackage{dsfont}
\usepackage{relsize}
\usepackage{tikz}	
\usetikzlibrary
{
	arrows,
	calc,
	through,
	shapes.misc,	
	shapes.arrows,
	chains,
	matrix,
	intersections,
	positioning,	
	scopes,
	mindmap,
	shadings,
	shapes,
	backgrounds,
	decorations.text,
	decorations.markings,
	decorations.pathmorphing,	
}
\begin {document}
  \newcommand {\nc} {\newcommand}
  \nc {\beq} {\begin{eqnarray}}
  \nc {\eeq} {\nonumber \end{eqnarray}}
  \nc {\eeqn}[1] {\label {#1} \end{eqnarray}}
  \nc {\eol} {\nonumber \\}
  \nc {\eoln}[1] {\label {#1} \\}
  \nc {\ve} [1] {\mbox{\boldmath $#1$}}
  \nc {\ves} [1] {\mbox{\boldmath ${\scriptstyle #1}$}}
  \nc {\mrm} [1] {\mathrm{#1}}
  \nc {\half} {\mbox{$\frac{1}{2}$}}
  \nc {\thal} {\mbox{$\frac{3}{2}$}}
  \nc {\fial} {\mbox{$\frac{5}{2}$}}
  \nc {\la} {\mbox{$\langle$}}
  \nc {\ra} {\mbox{$\rangle$}}
  \nc {\etal} {\emph{et al.}}
  \nc {\eq} [1] {(\ref{#1})}
  \nc {\Eq} [1] {Eq.~(\ref{#1})}
  \nc {\Ref} [1] {Ref.~\cite{#1}}
  \nc {\Refc} [2] {Refs.~\cite[#1]{#2}}
  \nc {\Sec} [1] {Sec.~\ref{#1}}
  \nc {\chap} [1] {Chapter~\ref{#1}}
  \nc {\anx} [1] {Appendix~\ref{#1}}
  \nc {\tbl} [1] {Table~\ref{#1}}
  \nc {\Fig} [1] {Fig.~\ref{#1}}
  \nc {\ex} [1] {$^{#1}$}
  \nc {\Sch} {Schr\"odinger }
  \nc {\flim} [2] {\mathop{\longrightarrow}\limits_{{#1}\rightarrow{#2}}}
  \nc {\textdegr}{$^{\circ}$}
  \nc {\inred} [1]{\textcolor{red}{#1}}
  \nc {\inblue} [1]{\textcolor{blue}{#1}}
  \nc {\IR} [1]{\textcolor{red}{#1}}
  \nc {\IB} [1]{\textcolor{blue}{#1}}
  \nc{\pderiv}[2]{\cfrac{\partial #1}{\partial #2}}
  \nc{\deriv}[2]{\cfrac{d#1}{d#2}}
\title{Sensitivity of one-neutron knockout to the nuclear structure of halo nuclei}
\author{C.~Hebborn}
\email{chloe.hebborn@ulb.ac.be}
\affiliation{Physique Nucl\'eaire et Physique Quantique (CP 229), Universit\'e libre de Bruxelles (ULB), B-1050 Brussels}
\author{P.~Capel}
\email{pcapel@uni-mainz.de}
\affiliation{Institut f\"ur Kernphysik, Johannes Gutenberg-Universit\"at Mainz, D-55099 Mainz}
\affiliation{Physique Nucl\' eaire et Physique Quantique (CP 229), Universit\'e libre de Bruxelles (ULB), B-1050 Brussels}
\date{\today}
\begin{abstract}
\begin{description}
\item[Background] 
Information about the structure of halo nuclei are often inferred from one-neutron knockout reactions.
Typically the parallel-momentum distribution of the remaining core is measured after a high-energy collision of the exotic projectile with a light target.
\item[Purpose]
{We study how the structure of halo nuclei affects knockout observables considering an eikonal model of reaction.} 
\item[Method] To evaluate the sensitivity of both the diffractive and stripping parallel-momentum distributions to the structure of halo nuclei, we consider several descriptions of the projectile within a halo effective field theory.
We consider the case of $^{11}\rm Be$, the archetypical one-neutron halo nucleus, impinging on $^{12}\rm C$ at 68~MeV/nucleon, which are usual experimental conditions for such measurements.
{The low-energy constants of the description of $^{11}$Be are fitted to experimental data as well as {to} predictions of an \emph{ab initio} nuclear-structure model.}
\item[Results] One-neutron knockout reaction is confirmed to be purely peripheral, the parallel-momentum distribution of the remaining core is only sensitive to the asymptotics of the ground-state wavefunction and not to its norm. The presence of an excited state in the projectile spectrum reduces  the amplitude of the breakup cross section; {the corresponding probability flux is transferred to the inelastic-scattering channel}.
Although the presence of a resonance in the core-neutron continuum significantly affects the energy distribution, it has no impact on the parallel-momentum distribution.
\item[Conclusions] One-neutron knockout cross section can be used to infer information about the tail of the ground-state wavefunction, viz. its asymptotic normalization coefficient (ANC).
{The independence of the parallel-momentum distribution on the continuum description makes the extraction of the ANC from this observable very reliable.}
\end{description}

\end{abstract}
\maketitle
%


\section{Introduction}\label{Introduction}
Halos are very exotic nuclear structures observed far from stability, close to the driplines.
Compared to stable nuclei, halo nuclei exhibit a very large matter radius~\cite{T96}.
This unusual size results from the low separation energy for one or two nucleons observed in these nuclei.
Thanks to that lose binding, the valence nucleons can tunnel far into the classically forbidden region and exhibit a high probability of presence at a large distance from the other nucleons.
They thus form a sort of diffuse halo around a compact core~\cite{HJ87}.
These structures challenge the usual description of the nucleus, which sees the nucleons pilling up and forming compact objects.
It is therefore important to better understand how they form in order to improve our knowledge on {the} nuclear structure within the entire nuclear chart.
Because of their strongly clusterized structure, halo nuclei are usually described as few-body objects: a compact core, which contains most of the nucleons, to which one or two nucleons are loosely bound.
Archetypical halo nuclei are $^{11}$Be, seen as a $^{10}$Be core with one neutron in its halo, and $^{11}$Li, seen as a $^{9}$Li core with a two-neutron halo.

Being very short-lived, halo nuclei cannot be probed with usual spectroscopic methods, but have to be studied through indirect techniques, such as reactions.
For example, elastic-scattering data provide information about the size of the nucleus~\cite{Dip09l,Dip12}.
Since they are very sensitive to the single-particle structure of nuclei, transfer reactions are particularly well suited to study halo nuclei \cite{Sch12,Sch13}.
In breakup reactions the core-halo structure dissociates through its interaction with a target, hence revealing the cluster structure of the nucleus \cite{P03,Fetal04}.
Experimentally, breakup reactions of halo nuclei are of great interest, because the cross sections are large thanks to the low binding energy of the halo nucleons.
In this work, we present a theoretical analysis of such reactions involving one-neutron halo nuclei, like $^{11}$Be.
In particular, we focus on inclusive breakup, also called one-neutron knockout~\cite{HT03}.
In these measurements, only the core of the nucleus is detected \cite{Aetal00,Tetal02PRC,Setal04,Fang04}.
Contrary to exclusive measurements, in which both the core and the halo nucleon(s) are measured in coincidence \cite{P03,Fetal04}, inclusive measurements exhibit a much higher statistics and hence are often favored for the low-intensity beams available at radioactive-ion beam facilities.

Theoretical models that describe the inclusive breakup of two-body projectiles have been developed in the eighties in Refs.~\cite{HM85,IAV85}.
The corresponding cross sections are obtained as the sum of the cross section for the diffractive---or elastic---breakup, in which the collision leads to the dissociation of the halo neutron from the core, and that of the stripping, where only the core survives the reaction and the neutron is absorbed by the target.
These models treat the remaining core as a spectator, which is merely scattered elastically off the target. As they occur at intermediate to high energies, these reactions are often analyzed within the eikonal model~\cite{G59,HT03,Tetal02PRC,Aetal00,Setal04,Fang04}. Recently, this framework has been extended to three-body projectiles~\cite{CFH17,Setal18}. 

The goal of this work is to determine the physics of one-neutron halo nuclei probed through inclusive breakup.
{For this, we describe the one-neutron halo projectile within a halo effective field theory (Halo-EFT~\cite{BHK02}, see \Ref{HJP17} for a recent review).}
This model exploits the clear separation of scales observed in halo nuclei, viz. the large size of the halo $R_{ \rm halo}$ compared to the compact size of the core $R_{ \rm core}$, to expand the projectile Hamiltonian upon the small parameter $\frac{R_{\rm core}}{R_{ \rm halo}}<1$.
This very systematic expansion enables us to identify the nuclear-structure observables, which affect most the reaction process.
Following Refs.~\cite{CPH18,MC19}, we apply this method up to next-to-leading order (NLO) to simulate the $^{10}$Be-$n$ interaction.
Here we focus on the collision of $^{11}$Be on $^{12}$C at 68~MeV/nucleon and study in detail the sensitivity to the description of $^{11}$Be of the parallel-momentum distribution of the $^{10}$Be core following the inclusive breakup of the projectile.

We begin by presenting the three-body model of the reaction in \Sec{Sec2}. In Sec.~\ref{Sec3a}, we provide the numerical inputs and the optical potentials considered in this study. Then, we {analyze} in Sec.~\ref{Sec3b}, the sensitivity of the parallel-momentum distribution of the remaining $^{10}\mathrm{Be}$ core to the $^{11}\mathrm{Be}$ ground-state wavefunction.  For this purpose, we {consider} various Halo-EFT potentials, generating different ground-state wavefunctions.
In Secs.~\ref{Sec3c} and~\ref{Sec3d}, we study the sensitivity of breakup observables to other features in the description of the projectile, {namely} the presence of an excited subthreshold bound state and the description of {the continuum.}
The conclusions drawn from these three analyses are summarized in Sec.~\ref{Conclusions}.

\section{Reaction Model}\label{Sec2}
We consider the knockout of a one-neutron halo nucleus projectile $P$ on a target $T$. As mentioned in the Introduction, halo nuclei exhibit strongly clusterized structures. Accordingly, we model one-neutron halo nuclei as two-body objects, composed of a spinless core $c$ and a loosely-bound neutron $n$.
The structure of the halo nucleus is thus described by the internal {single-particle} Hamiltonian
	\begin{equation}
h_{cn} = \frac{p^2 }{2 \mu_{cn}}  + V_{cn}(r).\label{eq1}
\end{equation}
where $\ve{p}$ and $r$ are, respectively, the $c$-$n$ relative momentum and distance, $\mu_{cn}$ is the $c$-$n$ reduced mass and $V_{cn}$ is an effective potential simulating the $c$-$n$ interaction. As aforementioned, halo nuclei are good candidates for EFT-expansion. In this work, we follow  Refs.~\cite{CPH18,MC19}: we simulate the $c$-$n$ interaction with a Halo-EFT potential, and we constrain its low-energy constants  with the experimental binding energy of the bound states  and with theoretical predictions provided by \textit{ab initio} calculations~\cite{Cetal16} (see Sec.~\ref{Sec3a}).

\begin{figure}
	\centering
	{\includegraphics[width=\linewidth]{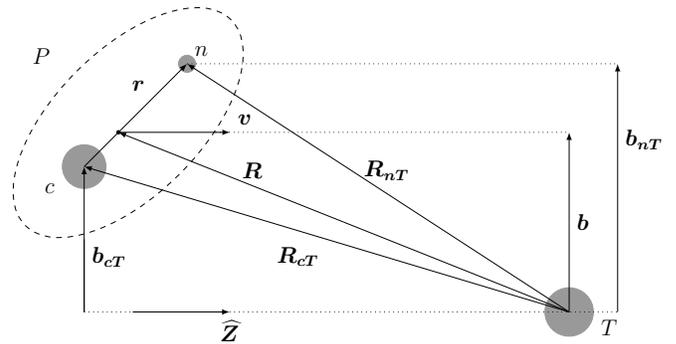}}
	\caption{\label{Fig3BodyCoordinates} Set of coordinates of the three-body model of the collision: the $c$-$n$ relative coordinate $\ve{r}$; the relative coordinate $\ve{R}$ between the projectile center-of-mass and the target and its component $\ve{b}$ transverse to the beam axis $\ve{\widehat Z}$, the $c$-$T$ and $n$-$T$ relative coordinates $\ve{R_{cT}}$ and $\ve{R_{nT}}$ with their transverse parts $\ve{b_{cT}}$ and $\ve{b_{nT}}$, respectively.}
\end{figure} 

The {single-particle} eigenstates  $\phi_{lJM}$ of $h_{cn}$, characterizing the $c$-$n$ relative  motion, are solutions of
	\begin{equation}
h_{cn}\,\phi_{lJM}(E,\ve{r})=E\,\phi_{lJM}(E,\ve{r}),\label{eq2}
\end{equation}
where $l$ is the orbital angular momentum of the $c$-$n$ system, $J$ is the total angular momentum, resulting from the composition of $l$ and the spin of the neutron $s$, and $M$ is its projection.  
These eigenstates can be   expressed from their radial part $u_{lJ}$, a spinor $\chi_s^{m_s}$ and spherical harmonics $Y_l^{m_l}$
	\begin{eqnarray}
	\phi_{lJM}(E,\ve{r})&=& \frac{u_{lJ}(E,r)}{r}\,[\chi_s \otimes Y_l (\ve{\hat{r}})]^{JM}.
	\end{eqnarray}
	
The eigenvalues $E$ can be positive or negative. The negatives energies $E_{nlJ}$ are discrete and correspond to bound states. These states are characterized by an additional quantum number, the number of nodes $n$ in the radial wavefunction. Asymptotically, their radial part behaves as
	\begin{equation}
		u_{nlJ}(E_{nlJ},r) \flim{r}{\infty}{b}_{nlJ}\,i \,\kappa_{nlJ}\,  h_l^{(1)}(i\kappa_{nlJ} r)\label{eq4}
	\end{equation}
{where ${b}_{nlJ}$ is the single-particle asymptotic normalization coefficient (SPANC), $\kappa_{nlJ}= \sqrt{2 \mu_{cn}|E_{nlJ}|/\hbar^2}$ and {$ h_l^{(1)}$ is a modified spherical Bessel function of the third kind~\cite{AS70}}.  In the actual structure of the nucleus, several single-particle configurations $lJ$ contribute to each  projectile state. The $c$-$n$ overlap wavefunction in the configuration $lJ$ is characterized by  the  asymptotic normalization constant 	$\mathcal{C}_{lJ}$ (ANC). One can relate the 	projectile ANC	$\mathcal{C}_{lJ}$   to the SPANC ${b}_{nlJ}$ through the spectroscopic factor $S_{nlJ}$ }
\begin{eqnarray}
{\mathcal{C}_{lJ}=\sqrt{S_{nlJ}}\,b_{nlJ}}.\label{eq4bis}
\end{eqnarray}
{When $S_{nlJ}=1$, the projectile is described by only one single-particle state and the normalization constants are equal.}

	The positive-energy part of the spectrum is continuous and describes the states in which the neutron is not bound to the core. These states are associated with the $c$-$n$ wave number $k= \sqrt{2 \mu_{cn}E/\hbar^2}$. Their radial components tend asymptotically to
	\begin{equation}
			u_{lJ} (E,r)\flim{r}{\infty}\cos{\left[\delta_{lJ}(E)\right]}\, k r \,j_l(kr) + \sin\left[\delta_{lJ}(E)\right]\, kr\, n_l(kr),\label{eq5}
	\end{equation}
	where $\delta_{lJ}$ is the phase shift and $j_l$ and $n_l$ are the spherical Bessel functions of the first and second kind, respectively~\cite{AS70}.

As usual in reaction theory, we neglect the structure of the target and simulate its interaction with the projectile constituents $c$ and $n$ by local optical potentials $V_{cT}$ and $V_{nT}$, respectively \cite{BC12}.
Within this framework, the $P$-$T$ relative motion is described by  the three-body wavefunction $\Psi$, solution of the \Sch equation
	\begin{eqnarray}
	\lefteqn{\left[\frac{P^2}{2\mu}+h_{cn}+V_{cT}(R_{cT})+V_{nT}(R_{nT})\right]\Psi(\ve{R},\ve{r})=}\nonumber\\
	&\hspace{5.5cm}&E_{\rm tot}\ \Psi(\ve{R},\ve{r}), \label{eq6}
	\end{eqnarray}
	where $\ve{P}$ and $\ve{R}$ are respectively the $P$-$T$ relative momentum and coordinate (see the coordinate system illustrated in Fig.~\ref{Fig3BodyCoordinates})  and $\mu$ is the $P$-$T$ reduced mass. {This equation is solved with the initial condition that the projectile is in its ground state $\phi_{n_0l_0J_0M_0}$ and is impinging on the target along the beam direction, that we choose to be the $Z$ axis, i.e.,
	$	\Psi^{(M_0)}(\ve{R},\ve{r})\flim{Z}{- \infty}\exp(iKZ+\cdots)\ \phi_{n_0l_0J_0M_0} (E_{n_0l_0J_0},\ve{r})$. The total energy of the system is therefore fixed  by the sum of the projectile ground-state energy  and the initial kinetic energy $E_{\rm tot}= E_{n_0l_0J_0}+ \frac{\hbar^2 K^2}{2\mu}$.}

{In this work, we solve this three-body \Sch equation within the eikonal model~\cite{G59,BC12}. To solve the divergence of the breakup matrix element due to the Coulomb interaction, we use the Coulomb-corrected eikonal model (CCE), presented in Refs.~\cite{MBB03,CBS08}. The expressions of the stripping and the diffractive-breakup cross sections can be found respectively in Refs.~\cite{HM85,KBE96} and Ref.~\cite{CBS08}.}

\section{Results}\label{Sec3}
\subsection{Numerical inputs and two-body interactions}\label{Sec3a}
To conduct this sensitivity analysis, we consider  the one-neutron knockout of  $^{11}\mathrm{Be}$ on a $^{12}\mathrm{C}$ target at 68~MeV/nucleon. Within the single-particle model presented in Sec.~\ref{Sec2}, we describe the $1/2^+$ ground state of the one-neutron halo nucleus $^{11}\mathrm{Be}$ as an inert  $^{10}\mathrm{Be}$ core, assumed to be in its $0^+$ ground state, to which an $s$-wave valence neutron is bound by 0.504~MeV. We follow Refs.~\cite{CPH18,MC19} and we simulate the $^{10}\mathrm{Be}$-${n}$ interaction with a Halo-EFT potential built with $\frac{R_{\rm core}}{R_{ \rm halo}}\sim 1/3$ as expansion parameter.
Accordingly, we consider for $V_{cn}$ in \Eq{eq1} purely contact interactions and their derivatives, which we regulate by Gaussians to obtain numerically tractable potentials.
As in Refs.~\cite{CPH18,MC19}, we truncate this expansion at the NLO and parametrize the potential per partial wave $lJ$ in the following way
\begin{equation}
V_{cn}^{lJ} (r)=V_{0}^{lJ} e^{-\frac{r^2}{2r_0^2}} + V_{2}^{lJ} r^2e^{-\frac{r^2}{2r_0^2}},\label{eq8}
	\end{equation}
where $V_0^{lJ}$ and $V_2^{lJ}$ are adjustable parameters, which can be fitted in each partial wave to reproduce experimental data or predictions from microscopic models.
The range of the Gaussians $r_0$ is an unfitted parameter, which can be varied to estimate the sensitivity of our calculations to the short-range physics of the projectile.

At NLO,  the two adjustable parameters, $V_0^{lJ}$ and $V_2^{lJ}$, have to be constrained in the $s$  and $p$ waves. 
In the $s1/2$ and $p1/2$ partial waves, we fit them to reproduce the experimental binding energies of the $1/2^+$ ($E_{1/2^+}=-0.504$~MeV) and $1/2^-$ ($E_{1/2^-}=-0.184$~MeV) bound states of $^{11}$Be.
These bound states are described by the single-particle states $1s1/2$ and $0p1/2$, respectively, with unit spectroscopic factors.
Halo-EFT potentials are also adjusted to the ANC of these states [see Eqs.~\eqref{eq4}--\eqref{eq4bis}] predicted by the \textit{ab initio} calculations of Calci \etal~\cite{Cetal16}: ${b}_{1s1/2}=\mathcal{C}_{s1/2}=0.786$~fm$^{-1/2}$ and ${b}_{0p1/2}=\mathcal{C}_{p1/2}=0.129$~fm$^{-1/2}$. 
We do not put any interaction in the $p3/2$ wave since the $p3/2$ phase shift predicted by  Calci~\etal\ is approximatively zero at low energy $E$.

To test the influence of the $1s1/2$ ground state on our reaction calculations, we generate various $s1/2$ Halo-EFT potentials. 
We consider two Gaussian ranges $r_0=1.2$~fm and 2.0~fm.
{Then, since the \textit{ab initio} calculations predict a spectroscopic factor $S_{1s1/2}=0.9$ for the $1s1/2$ configuration~\cite{Cetal16}, we also fit the potentials to reproduce a wavefunction with the same ANC when its norm is reduced to $\sqrt{0.9}$, i.e., ${b}_{1s1/2}=0.829$~fm$^{-1/2}$[$=0.786/\sqrt{0.9}$~fm$^{-1/2}$, see Eq.~\eqref{eq4bis}].}
The parameters $V_0^{s1/2}$ and $V_2^{s1/2}$ obtained from these different fits are displayed in Table~\ref{Tab11BePot} alongside the resulting eigenenergies and {SPANCs}.
The  $1s1/2$ wavefunctions generated from these potentials are plotted in Fig.~\ref{FigPeriph}(a).

{Similarly, in the $p1/2$ partial wave, we have considered the same two ranges for the Gaussian potential. However, since our calculations are insensitive to the choice of $r_0$ in this partial wave, we limit the results we display in this article to those obtained solely with $r_0=1.2$~fm, whose low-energy constants are displayed in the last line of Table~\ref{Tab11BePot}.}

\begin{table}
	\begin{tabular}{c|c|cc|cc}
		&	$r_0$ & $V_0^{lJ}$  & $V_2^{lJ}$ &$E_{nlJ}$ & ${{b}_{nlJ}}$\\
		&[fm] & [MeV]&[MeV]&[MeV]&[fm$^{-1/2}$]\\ \hline\hline
		\multirow{3}{*}{1$s_{1/2}$}&1.2 &-50.375 &-45 &-0.504& 0.786\\
		&2 &-80.54 &2.97 &-0.504& 0.786\\
		&1.2 &86.03 &-108.62 &-0.504&  0.829\\\hline
		0$p_{1/2}$&1.2& -96.956& 0&-0.184 & 0.129\\
	\end{tabular}
	\caption{Depths of the Halo-EFT potential~\eqref{eq8} at  NLO used to simulate the $^{10}\mathrm{Be}$-$n$ interaction in the $s1/2$ and $p1/2$ partial waves. The depths are fitted to the experimental binding energy and the  ANC predicted by Calci \etal~\cite{Cetal16}.}\label{Tab11BePot}
\end{table}

The $P$-$T$ nuclear interactions are simulated by Woods-Saxon optical potentials 
	\begin{eqnarray}
			V (R) &=& -V_R\,f_{\rm WS}(R,R_R,a_R) -i\,W_I\,f_{\rm WS}(R,R_I,a_I) \nonumber \\
	&& +i\,4 a_D W_D \deriv{}{R} f_{\rm WS}(R,R_D,a_D), \label{eq13}
	\end{eqnarray}
	where
	\begin{equation}
			f_{\rm WS}(R,R_X,a_X) = \frac{1}{1 +e^{\frac{R-R_X}{a_X}}}.\label{eq14}
	\end{equation}
 	For the  $^{10}\mathrm{Be}$-$^{12}\mathrm{C}$ interaction, we use the parameters of  Ref.~\cite{AKTB97}, which are consistent with data for the $^{10}\mathrm{Be}$-$^{12}\mathrm{C}$ elastic scattering at 59~MeV/nucleon.  The Coulomb interaction is simulated by a potential generated by a uniformly charged sphere of radius $R_C=1.2\,(10^{1/3}+12^{1/3})$~fm. The $n$-$^{12}\mathrm{C}$ interaction is modeled by the potential developed in Ref.~\cite{W18}, fitted to  elastic scattering data of a nucleon off a nucleus with $A\leq 13$ at energies between 65~MeV and 75~MeV~\footnote{We obtain similar results with other realistic $n$-$^{12}\mathrm{C}$ potentials given in Refs.~\cite{W09,CK80}.}. For both potentials, we neglect any energy dependence. The parameters of the two optical potentials used in this study are listed in Table~\ref{TabOptPot}.
	
	For all the computations, we use the following model space: the $^{10}\mathrm{Be}$-$n$ continuum is described up to the $c$-$n$ orbital angular momentum $l_{\rm max}=10$ and a mesh in impact parameter up to 100~fm, with a step of 0.25~fm up to 30~fm and of 2~fm beyond.
	All the parallel-momentum distributions of the diffractive breakup are integrated up to $k_{\rm max}=1.5$~fm$^{-1}$, which corresponds to $E_{\rm max}=51.3$~MeV. {In this article,  these distributions are centered at the projectile center-of-mass parallel-momentum.} The total breakup cross sections are obtained by integrating the energy distribution.  The total uncertainties made on these computations are of the order of $0.6\%$.

\begin{table*}
	\begin{tabular}{c|ccccccccc|c}
		& $V_R$ [MeV] & $R_R$ [fm] & $a_R$ [fm] & $W_I$ [MeV]& $R_I$ [fm]&$a_I$ [fm] &$W_D$ [MeV] & $R_D$ [fm] &$a_D$ [fm] &Ref.\\ \hline \hline
		{$^{10}\mathrm{Be}$-$^{12}\mathrm{C}$}&123.0& 3.33& 0.8&  65.0& 3.47&  0.8& &&& {\cite{AKTB97}}\\
		$n$-$^{12}\mathrm{C}$&31.5& 2.65 &0.65& 5.25 & 2.65&0.65&7.66 &3.24& 0.178 &{\cite{W18}}\\
	\end{tabular}	
	\caption{Parameters of the Woods-Saxon optical potentials~\eqref{eq13}--\eqref{eq14} used to simulate the  $^{10}\mathrm{Be}$-$^{12}\mathrm{C}$ and $n$-$^{12}\mathrm{C}$ interactions. }
	\label{TabOptPot}
\end{table*}

\subsection{Sensitivity to the ground-state wavefunction}\label{Sec3b}

\begin{figure*}
	\center
	{\includegraphics[width=0.48\linewidth]{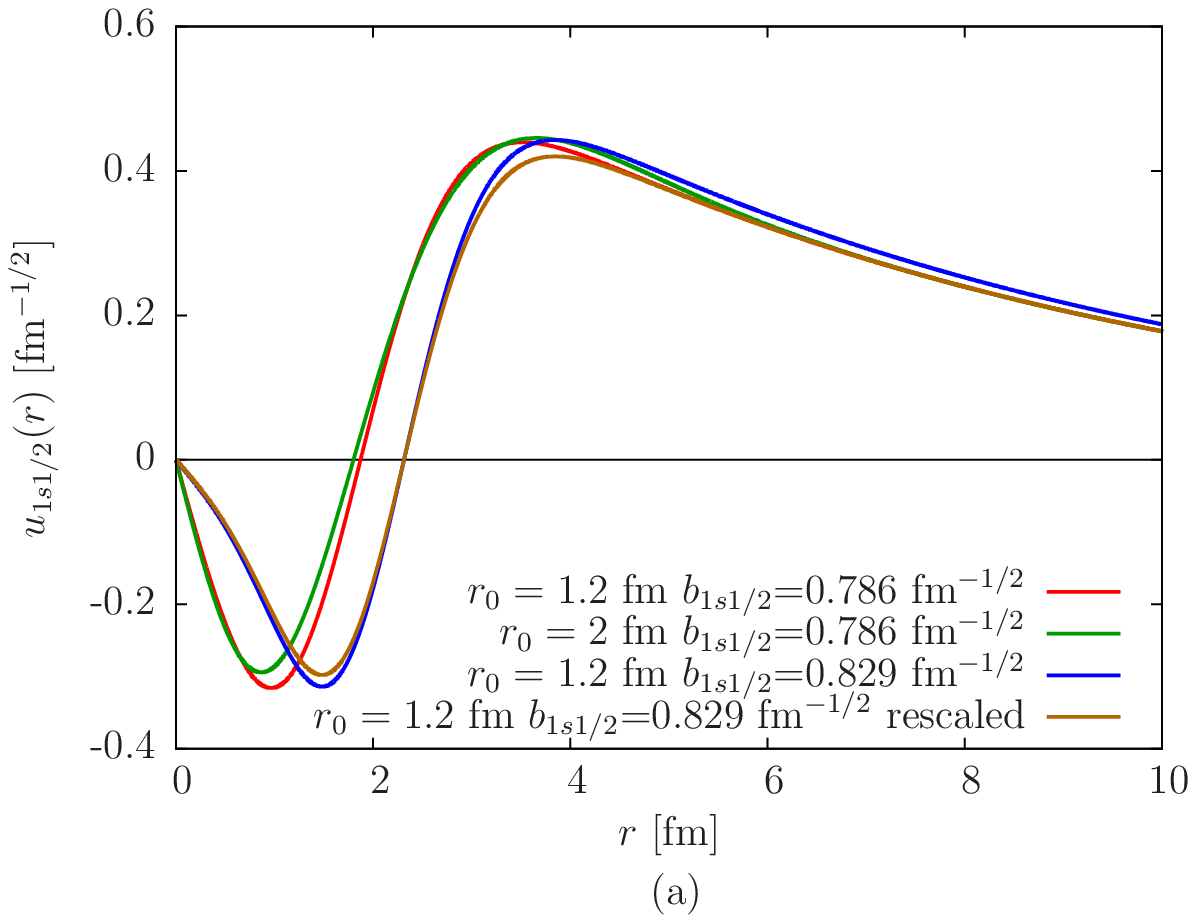}}
	\hspace{0.3cm}
	{	\includegraphics[width=0.48\linewidth]{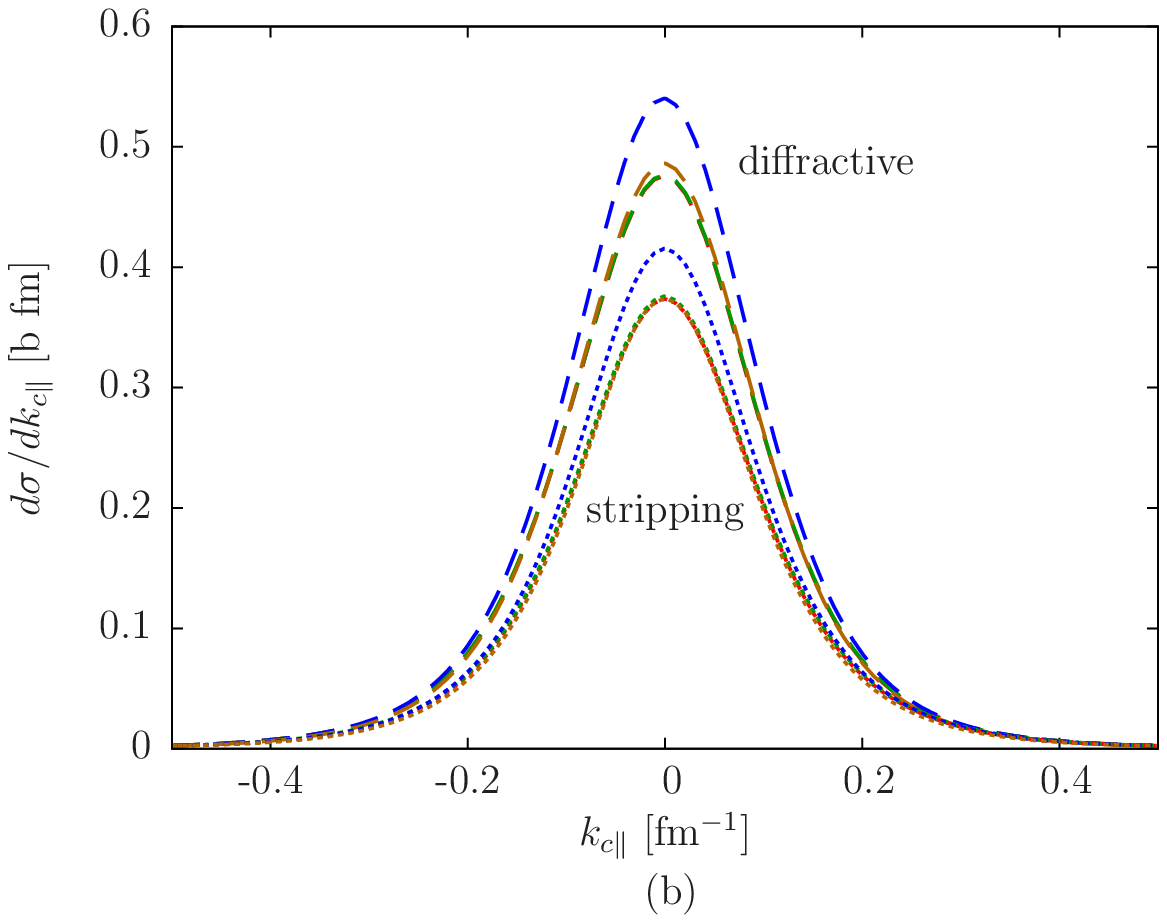}}
	\caption{(a) Radial wavefunctions of the $1s1/2$  ground state of $^{11}\mathrm{Be}$, obtained with potentials {reproducing ${b}_{1s1/2}=0.786$~fm$^{-1/2}$ with a range $r_0=1.2$~fm (red line) and with $r_0=2$~fm (green line), reproducing  ${b}_{1s1/2}=0.829$~fm$^{-1/2}$ (blue lines) and rescaled  with $\frac{0.786}{0.829}$ (brown line).} (b) Parallel-momentum distribution of $^{10}\mathrm{Be}$ resulting from the diffractive breakup (dashed lines) and the stripping (dotted lines) of $^{11}\mathrm{Be}$ on $^{12}\mathrm{C}$ at 68~MeV/nucleon. The colors used in the cross sections correspond to those of the ground-state wavefunctions of subfigure (a).  	}\label{FigPeriph}
\end{figure*}

As detailed in \Sec{Sec3a}, we have generated different $^{10}\mathrm{Be}$-$n$ potentials leading to various $1s1/2$ ground-state wavefunctions. 
The corresponding ground-state wavefunctions are plotted in Fig.~\ref{FigPeriph}(a).
One can see that   the two wavefunctions obtained with different ranges ($r_0=1.2$~fm in red line and $r_0=2$~fm in green line) differ slightly below $6$~fm but exhibit identical asymptotics.  The ground-state wavefunction reproducing a larger {SPANC} (blue lines) has larger asymptotics and a very different short-range behavior.
To determine if the breakup process is sensitive only to the asymptotics, we normalize this new wavefunction to the spectroscopic factor ${0.9}$ predicted by Calci \etal\ \cite{Cetal16}. By construction, this new wavefunction (brown lines) exhibits the same asymptotics as the previous ones while being very different below $r\approx 4$~fm. 

The corresponding parallel-momentum distributions of $^{10}\mathrm{Be}$ for the diffractive breakup (dashed lines) and stripping (dotted lines)  of $^{11}\mathrm{Be}$ on $^{12}\mathrm{C}$ at 68~MeV/nucleon are plotted in Fig.~\ref{FigPeriph}(b).
 The two cross sections obtained with the potentials fitted with $r_0=1.2$~fm and $r_0=2$~fm (red and green lines, respectively) superimpose perfectly for both the stripping  and the diffractive processes. 
{This confirms the results of Refs.~\cite{H96,BH02,CN07} which show that these observables are not sensitive to changes in the ground-state wavefunction at small distance $r$.}
When the reaction is computed with the ground-state wavefunction fitted to the larger {SPANC} (blue lines), we observe an increase of about 10\% in both cross sections.
After scaling that initial wavefunction to the $0.9$ spectroscopic factor predicted by Calci \etal\ \cite{Cetal16}, we obtain cross sections nearly identical to the previous ones (brown lines).
We can therefore conclude that, as the exclusive breakup~\cite{CN07}, the inclusive breakup of one-neutron halo nuclei is purely peripheral, in the sense that it is sensitive only to the tail of the initial ground-state wavefunction. {This is reminiscent of the result obtained by Hansen using a simple geometric model, where the stripping cross sections are shown to be proportional to the square of the {SPANC}~\cite{H96}, and to the confirmation of this result within an eikonal framework~\cite{BH02}.}
While all three calculations provide identical stripping cross sections, we observe a tiny difference in the diffractive part.
Further analyses have shown that this comes from the contributions at high $^{10}$Be-$n$ relative energies ($E>30$~MeV), where the process starts to be slightly more sensitive to the projectile radial wavefunction at small distances, viz. $r<4$~fm.

This analysis confirms that the knockout process is a peripheral reaction.
Therefore, information about the internal part of the wavefunction cannot be reliably inferred from such measurements.
This is in particular true for the norm of the overlap wavefunction, i.e., the spectroscopic factor.
Since calculations performed with two wavefunctions that exhibit different norms but the same ANC provide {nearly identical} results, it is not clear how accurate the spectroscopic factors extracted from knockout measurements are.
However, what is clear from this analysis, is that the parallel-momentum distributions for both diffractive breakup and stripping, are sensitive to the asymptotics of the ground-state wavefunction.
It suggests that these observables would be good candidates to extract accurately the ANC of the wavefunction of halo nuclei, as done in Ref.~\cite{TetalTAM02}.
To confirm this, we analyze in the next sections the sensitivity of these observables to other features of the projectile description, viz.\ the presence of an excited subthreshold bound state (\Sec{Sec3c}) and the description of the projectile continuum (\Sec{Sec3d}).

\subsection{Influence of excited subthreshold states}\label{Sec3c}

We now investigate how the presence of the $1/2^-$ excited state in the $^{11}\rm Be$ description affects knockout observables. Due to the form of the stripping cross section, which depends only on the ground state \cite{HM85,KBE96}, we restrict this study to the sole diffractive  breakup. As previously explained, we describe this $1/2^-$ bound state as a $0p1/2$ single-particle state, using the Halo-EFT $^{10}$Be-$n$ potential \eq{eq8} with the parameters listed in the last line of Table~\ref{Tab11BePot}.
The presence of that subthreshold state significantly affects the low-energy continuum in the $p1/2$ partial wave \cite{CN06,SCB10,CPH18}, which itself affects the calculation of breakup cross sections at low energy \cite{CN06,CPH18}.
We therefore expect to see some influence of that state in the diffractive component of the parallel-momentum distribution of the $^{10}$Be core following the breakup of $^{11}$Be.
To investigate this in detail, we consider two $^{10}$Be-$n$ interactions in that partial wave.
In addition to the $V_{cn}^{p1/2}$ potential described in \Sec{Sec3a}, we consider no interaction at all, hence without considering the $1/2^-$ excited state of $^{11}$Be and describing the $^{10}$Be-$n$ motion in the $p1/2$ continuum by mere plane waves.

\begin{figure}
	\center
	{\includegraphics[width=\linewidth]{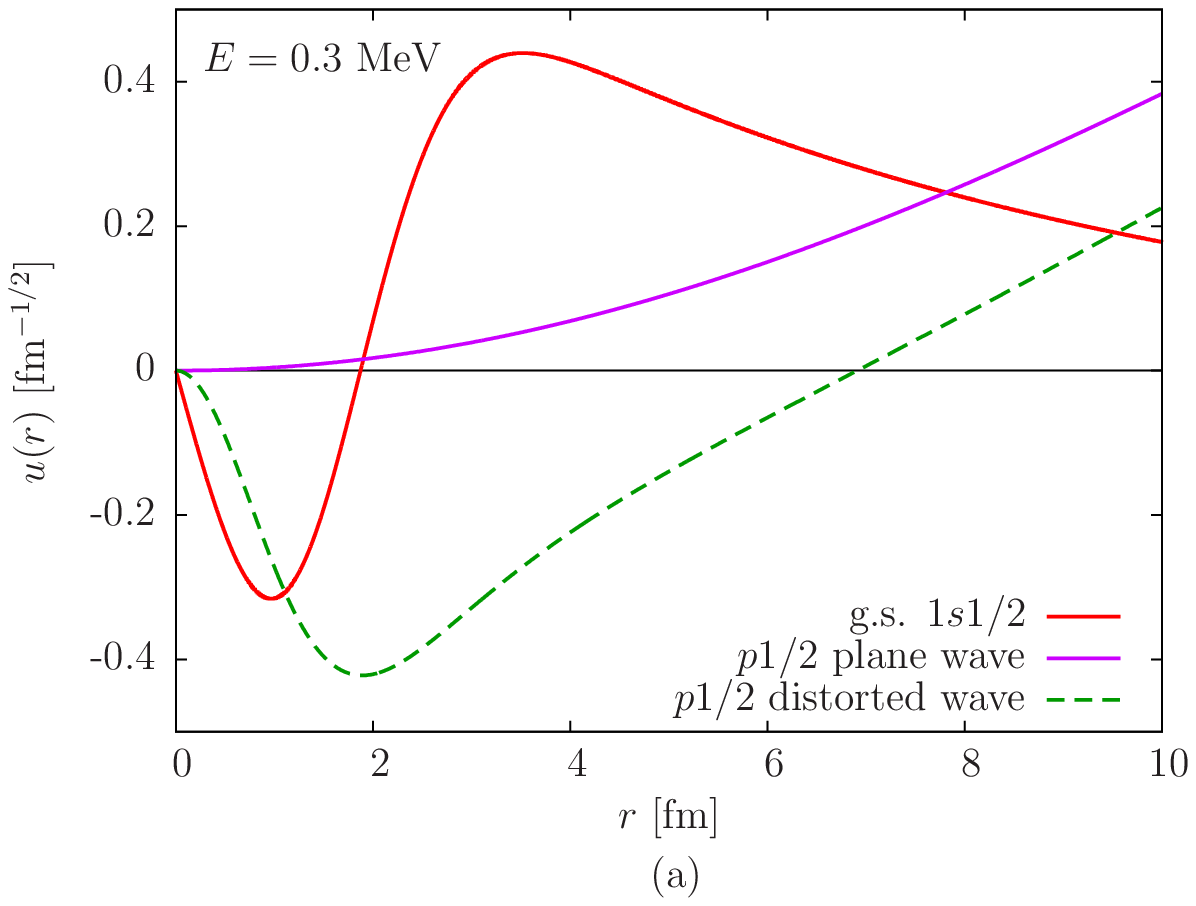}}
	{\includegraphics[width=\linewidth]{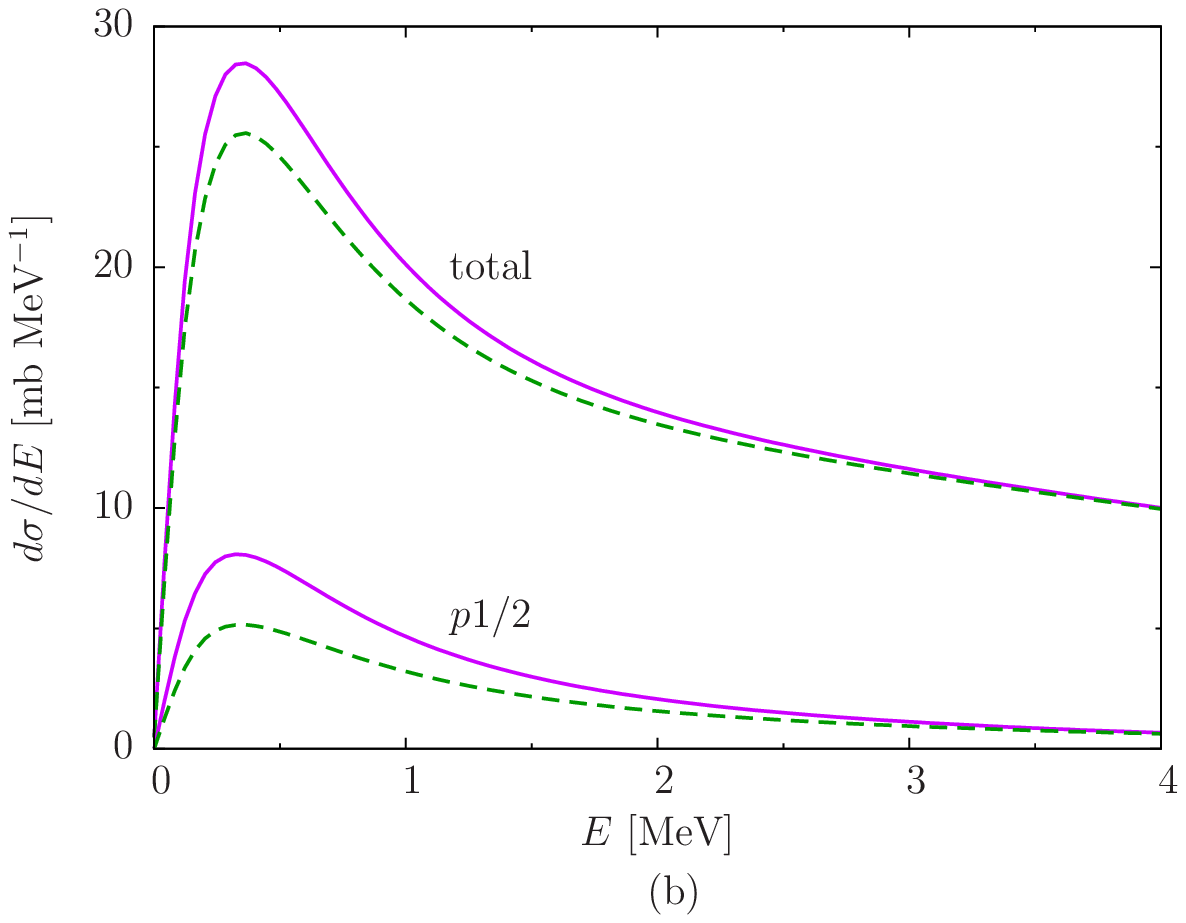}}
	{	\includegraphics[width=\linewidth]{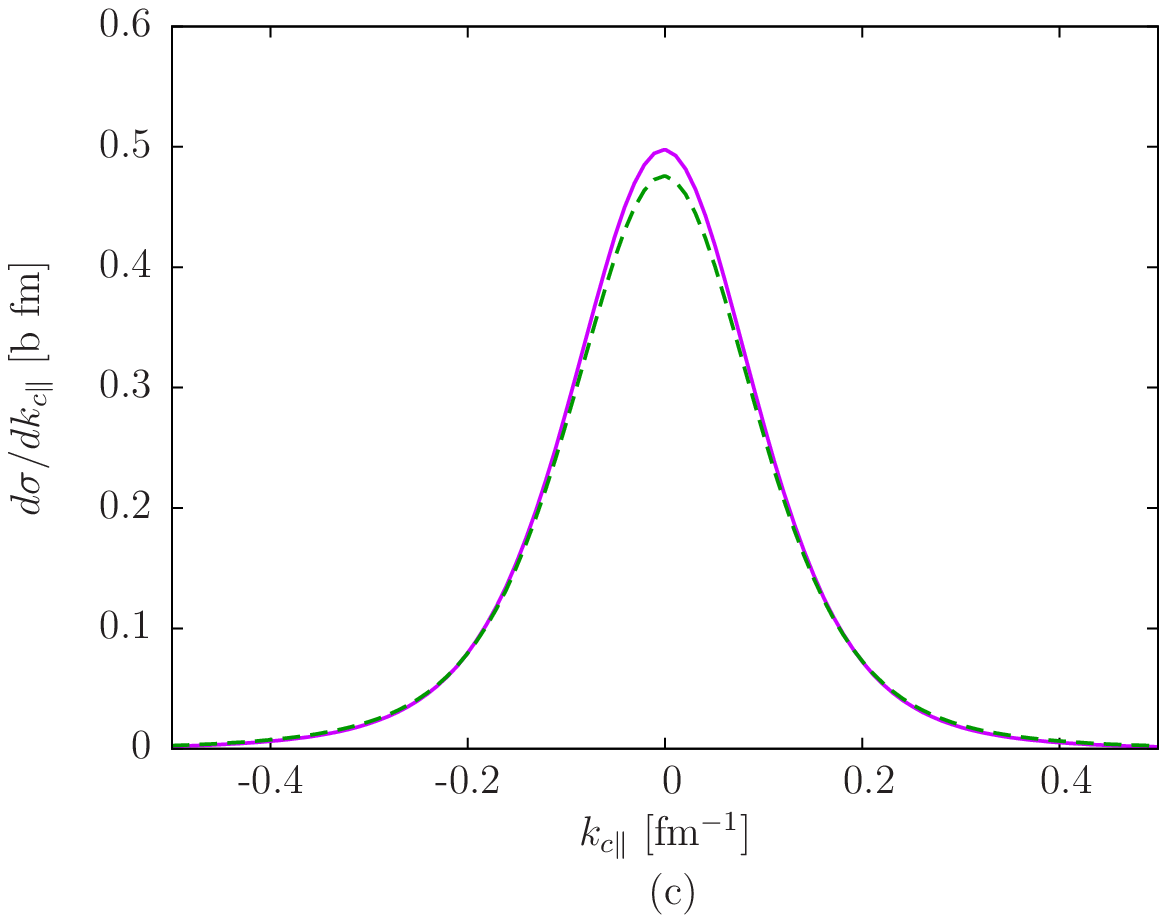}}
	
	\caption{Influence of the presence of a subthreshold bound state in the projectile spectrum on breakup observables for $^{11}\mathrm{Be}$ on $^{12}\mathrm{C}$ at 68~MeV/nucleon.
(a) Radial wavefunctions for different $c$-$n$ interactions in the $p1/2$ waves at $E=0.3$~MeV in the $^{10}\mathrm{Be}$-$n$ continuum and of the $1s1/2$ ground-state. 
Diffractive-breakup cross section as a function of (b) the $^{10}\mathrm{Be}$-$n$ relative energy $E$ {(total and $p1/2$ contribution)} and (c) the parallel-momentum distribution of $^{10}\mathrm{Be}$.}\label{FigBS}
\end{figure}

Figure~\ref{FigBS}(a) shows the radial wavefunctions for the $p1/2$ waves in the continuum at $E=0.3$~MeV.
The distorted wave obtained with the Halo-EFT potential is shown as a dashed green line, while the plane wave is displayed in solid magenta line.
The presence of the $1/2^-$ bound state affects the distorted waves in two ways.
It induces a node at $r\sim6.5$~fm and it produces a non-zero phase shift. {For comparison, the radial wavefunction $u_{1s1/2}$ of the $^{11 }\rm Be$ ground state is displayed as well (solid red line).}

The cross section for the diffractive breakup of $^{11}\rm Be$ on $^{12}\rm C$ at 68~MeV/nucleon is displayed in Fig.~\ref{FigBS}(b) as a function of the $^{10}\rm Be$-$n$ relative energy $E$ and in Fig.~\ref{FigBS}(c) as a function of the $^{10}\mathrm{Be}$ parallel momentum. Figure~\ref{FigBS}(b) shows also the contribution of the $p1/2$ partial wave separately (lower set of curves).
The major effect of the presence of the $1/2^-$ bound state in the description of the projectile is a reduction of the $p1/2$ diffractive breakup, mostly at low energies in the continuum [see \Fig{FigBS}(b)].
This also leads to a drop, albeit less significant, of the parallel-momentum distribution by about $4.4$\% [see \Fig{FigBS}(c)].
Interestingly, only the $p1/2$ contribution is affected by the presence of the $1/2^-$ state. 
This reduction is quantified by the cross sections provided in Table~\ref{TabBS}.
When we shift from the description of $^{11}$Be that includes both bound states (first column) to that where there is no interaction in the $p1/2$ partial wave (second column), the inelastic cross section $\sigma_{\rm inel}$ is practically entirely transferred to the breakup channel $\sigma^{\rm total}_{\rm bu}$, and, more precisely, to its sole $p1/2$ contribution $\sigma^{\rm p1/2}_{\rm bu}$.

This decrease in the cross section can be qualitatively explained by looking at the overlap of the radial wavefunction in the $p1/2$ continuum and the  $1s1/2$ ground state, which both appear in the matrix element for the breakup {[See Eq.~\eqref{eqA2} of Appendix~\ref{App1}]}. {Both the node at $r\sim 6.5$~fm and the phase shift introduced by the $p1/2$ interaction affect the breakup matrix element. To discriminate the impact of the node from the one of the phase shift, we have applied two different approaches. The first is to remove the $1/2^-$ state from the description of $^{11}$Be using phase-equivalent transformations of the $V_{cn}^{p1/2}$ potential through supersymmetry~\cite{S85,B87PRL,B87JPA}. These transformations conserve the phase shifts while eliminating the bound state, and hence the first node in the radial wavefunctions describing the $p1/2$ continuum. The second approach is to  use plane waves to describe the $p1/2$ continuum, that we orthogonalize to the $0p1/2$ wavefunction obtained from the $V_{cn}^{p1/2}$ of Table~\ref{Tab11BePot}. This generates a node at small distances in the continuum wavefunctions while keeping a nil phase shift. These two tests,  not presented here to keep the analysis concise, have shown that both the node and the phase shift contribute to that reduction.}

The same study performed  within the dynamical eikonal approximation (DEA)~\cite{BCG05}, where the adiabatic approximation is not considered, leads to identical results. Moreover, our conclusion remains unchanged when the excited bound state is in the $d$ wave, as would be the case for a $^{15}\rm C$ projectile, {another well known one-neutron halo nucleus}. {Therefore, such reactions at intermediate energies, where the dynamical effects are small,  conserve the probability flux within a partial wave, simply shifting that flux from the inelastic to the breakup channels. This effect has already been observed by Moro \etal\ in their theoretical analysis of the Coulomb-breakup measurement of $^{11}$Be performed at RIKEN \cite{Fetal04}.
	Including the $1/2^-$ bound excited state in the description of $^{11}$Be reduces the E1 strength to the continuum by an amount that is equal to the E1 strength for Coulomb excitation from the $1/2^+$ ground state to the $1/2^-$ excited state~\cite{Moro19}.}
	
{Theoretically, this transfer from the breakup to the inelastic channel is a consequence of the Hermicity of the $^{10} \rm Be$-$n$ Hamiltonian. Since the radial wavefunctions of the bound states  $u_{nlJ}$ and of the continuum $u_{lJ}(E)$ form an orthogonal basis in the subvectorial space defined by the partial wave $lJM$, we can  write the following closure relation}
	\begin{equation}
	{\sum_{n} \ket{u_{nlJ}}\bra{u_{nlJ}}+\frac{2}{\pi}\frac{\mu_{cn}}{\hbar^2 k} \int dE\ket{u_{lJ}(E)}\bra{u_{lJ}(E)}=\mathds{1}_{lJM},\label{eq11}}
	\end{equation}
	{where the sum runs over all the bound states in the partial wave $lJM$. By inserting this relation into the total diffractive-breakup cross section, we show in Appendix~\ref{App1} that {within the adiabatic approximation} the quantity $\sigma_{\rm  sum}^{lJM}$~\eqref{eqA6} only depends on the ground-state wavefunction and the $P$-$T$ interactions.}
{Accordingly, the sum of the total breakup and inelastic cross sections should be independent from the choice of $V_{cn}^{p1/2}$  (see the last line of Table~\ref{TabBS}). The small differences are due to numerical errors. }

\begin{table}
	\begin{tabular}[c]{c|c|c}
		&{1$s$1/2 + 0$p$1/2}& {1$s$1/2} + $p1/2$ plane wave  \\ \hline \hline
		$\sigma_{\rm bu}^{\rm total}$ [mb] &122.8 &126.1\\
		$\sigma_{\rm bu}^{p1/2}$ [mb] &10.2 &13.5 \\
		$\sigma_{\rm inel}$ [mb]   &3.6&0\\\hline
		$\sigma_{\rm bu}^{\rm total}+\sigma_{\rm inel}$ [mb]&126.4&126.1\\
	\end{tabular}
	\caption{Total diffractive breakup and inelastic cross sections of the collision  $^{11}\mathrm{Be}$ with $^{12}\mathrm{C}$ at 68~MeV/nucleon. They are obtained from computations considering both the $1/2^+$ ground state and the  $1/2^-$ excited state ($1s1/2$ + $0p1/2$) and when we set $V_{cn}^{p1/2}=0$ 
($1s1/2$ + $p1/2$ plane wave).}\label{TabBS}
\end{table}

In conclusion, the presence of an excited state changes non-negligibly  the shape and magnitude of the $c$-$n$ relative energy distribution for the diffractive breakup. The parallel-momentum distributions of the remaining core are affected in a smaller extent, i.e.,  less than 5\%  reduction of the peak amplitude. This reduction of the cross section  is  caused by both the node at short distance in the continuum wavefunctions and the non-zero phase shift introduced by the interaction.
{The amplitude loss in the diffractive  breakup goes to the inelastic-scattering channel, as already seen in Coulomb-breakup calculations by Moro \etal\ \cite{Moro19}. We have shown that this feature can be explained by the conservation of probability flux shared between the inelastic and breakup channels.}

\subsection{Sensitivity to the projectile's continuum}\label{Sec3d}

\begin{figure*}
	\center
	{\includegraphics[width=0.48\linewidth]{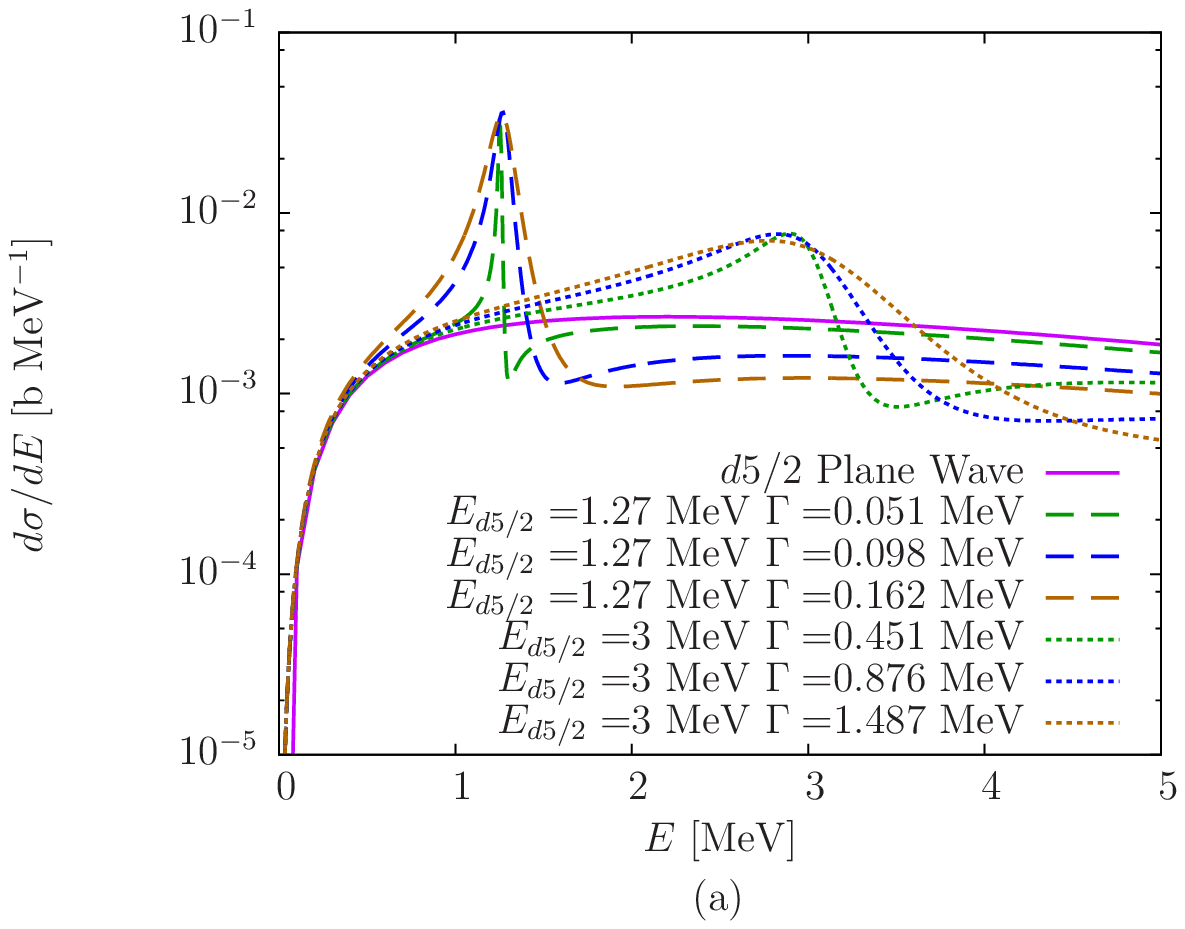}}
	\hspace{0.3cm}
	{	\includegraphics[width=0.48\linewidth]{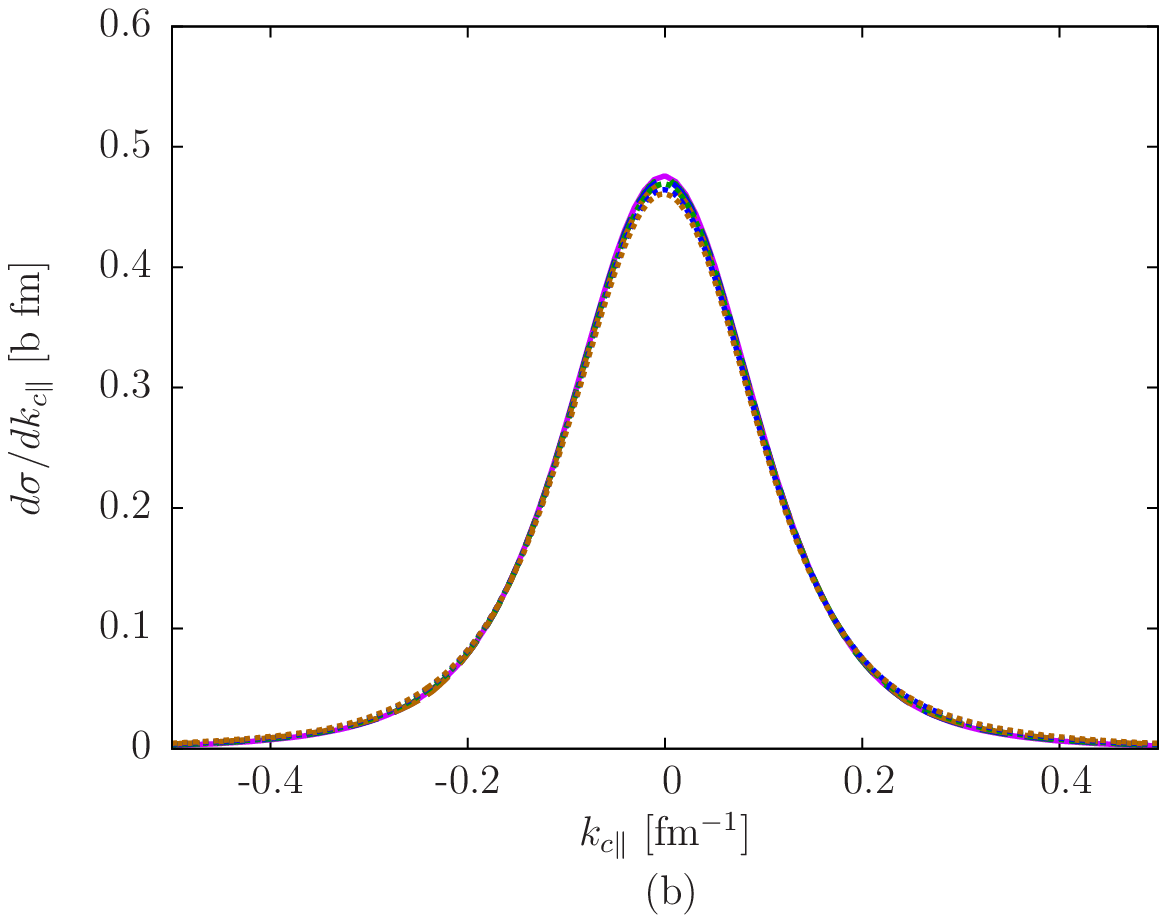}}
	\caption{Influence of a $d5/2$ resonance on breakup observables for $^{11}\mathrm{Be}$ on $^{12}\mathrm{C}$ at 68~MeV/nucleon.
(a) $d5/2$ contribution to the energy distribution and (b) breakup cross section as a function of the parallel-momentum of the remaining $^{10}\mathrm{Be}$. The solid magenta lines corresponds to the case where we consider no interaction in the $d5/2$-wave in the $^{10}\mathrm{Be}$-$n$ continuum, the dashed  and dotted lines correspond to cases in which a resonance is adjusted in the $d5/2$ continuum at $1.27$~MeV (actual $5/2^+$ state of $^{11}\rm Be$) and 3~MeV, respectively. The  colors vary with the width of these resonances.}\label{FigRes}
\end{figure*}
\begin{table*}
	\begin{tabular}[c]{c||c||c|c|c||c|c|c}
		& \multirow{2}{*}{$d5/2$ Plane Wave} & \multicolumn{3}{c||}{Res. $E_{d5/2}=1.27$ MeV}& \multicolumn{3}{c}{Res. $E_{d5/2}=3$ MeV}\\ 
		& & $\Gamma_{d5/2}=51$ keV&$\Gamma_{d5/2}=98$ keV &$\Gamma_{d5/2}=162$ keV & $\Gamma_{d5/2}=451$ keV&$\Gamma_{d5/2}=876$ keV &$\Gamma_{d5/2}=1487$ keV\\\hline\hline
		$\sigma_{\rm bu}^{\rm total}$ [mb] &122.8 &122.1 &122.6 &122.8 &122.8 &122.8 &122.6 \\
		$\sigma_{\rm bu}^{d5/2}$ [mb] &19.0 &18.3 &18.8 &19.0 &19.0 &19.0 & 18.8 \\
	\end{tabular}
	\caption{Integrated breakup cross sections of the collision  $^{11}\mathrm{Be}$ with $^{12}\mathrm{C}$ at 68~MeV/nucleon. They are obtained when we model both the $1/2^+$  ground state and the $1/2^-$ excited state,  with plane waves in $d5/2$ and with resonances at $E_{d5/2}=1.27$~MeV and at $E_{d5/2}=3$~MeV with different widths $\Gamma_{d5/2}$.}
	\label{TabRes}
\end{table*}

	 In this last part, we investigate how resonances in the $^{10}\mathrm{Be}$-$n$ continuum influence knockout observables. As in the previous section, we  study only the diffractive breakup because at the usual eikonal approximation the stripping cross section does not depend on the description of the continuum of the projectile~\cite{HM85,KBE96}. 
To do so, in addition to the plane waves used so far to describe the $d5/2$ continuum, we include a single-particle resonance in that partial wave at $E_{d5/2}=1.27$~MeV with a width of $\Gamma_{d5/2}=98$~keV, close to the experimental values of the physical $5/2^+$ resonance $E_{5/2^+}^{\rm exp}=1.274$~MeV and $\Gamma_{5/2^+}^{\rm exp}=100$~keV. This approach goes beyond the NLO of the Halo-EFT expansion, since we put an interaction in the $d$ wave.
To study in detail the impact of the continuum, we also consider resonances at the same energy with other widths, i.e., $\Gamma_{d5/2}=51$~keV and $\Gamma_{d5/2}=162$~keV, and  at a higher energy $E_{d5/2}=3$~MeV with various widths $\Gamma_{d5/2}=451$~keV, $\Gamma_{d5/2}=876$~keV and $\Gamma_{d5/2}=1487$~keV. To model these resonances, we vary the depths of the Gaussian potential~\eqref{eq8} in the sole $d5/2$ partial wave.

In  Fig.~\ref{FigRes}(a), we display the $d5/2$ contribution to the diffractive-breakup cross section for $^{11}\rm Be$ on $^{12}\rm C$ at 68~MeV/nucleon as a function of the $^{10}\rm Be$-$n$ relative energy. 
In addition to the different $d5/2$ partial-wave descriptions mentioned above, we also display the results obtained using plane waves in the $d5/2$ continuum (solid magenta lines).
As expected from the results of Refs.~\cite{Fetal04,CGB04}, the presence of a resonance in the $d$ continuum leads to a large peak in that contribution to the breakup energy distribution [see Fig.~\ref{FigRes}(a)].
The peak is centered on the resonance energy and its width is close to that of the resonance.
The various $V_{cn}^{d5/2}$ considered in this study thus lead to very different energy distributions.
However, each peak is followed by a depletion area resulting from destructive interferences caused by the phase shifts going over $\pi/2$. The range of this area is proportional to the peak width: sharper resonances have a steeper drop and tend more rapidly to the plane-wave computation after the resonance. When this distribution is integrated, these two effects compensate one another. {Following Eq.~\eqref{eqA6} of Appendix~\ref{App1}, the integrated breakup cross section, listed in Table~\ref{TabRes}, is conserved, even within a partial wave.
The small difference in the total breakup cross section, including the resonance at $E_{d5/2}=1.27$~MeV and with $\Gamma_{d5/2}=51$~keV, is due to uncertainties in the integration of the energy distribution, which is more tricky for such sharp variations.}

The corresponding parallel-momentum distributions of the remaining $^{10}\rm Be$ are displayed in Fig.~\ref{FigRes}(b).
Because they are obtained through the integration over the (transverse) momentum, they exhibit nearly no sensitivity to the choice of the interaction in the $d5/2$ partial wave.
These observables are therefore quite insensitive to the description of the continuum: the presence (or absence) of a resonance does not influence this inclusive observable. 
This is an interesting result since it means that, contrary to energy distributions, where resonances have a significant impact, a precise description of the continuum is not needed for an accurate computation of the parallel-momentum distributions.
Basically, using simple plane waves to describe the continuum is enough, but to the possible presence of a subthreshold bound state (see \Sec{Sec3c}).
This strongly reduces the uncertainty related to the description of the continuum that appears in energy distributions for diffractive breakup \cite{CN06,CN17}.

{This conclusion} is not specific for one partial wave, we have observed similar results when resonances are included within $p$ and $f$ waves. We have also conducted the same analysis within the DEA and the conclusions are identical, showing that the dynamics of the projectile does not affect this finding.
This independence from the continuum description shows that this observable is ideal to extract accurate information pertaining to the {asymptotics of} the initial ground state of the projectile, such as its ANC (see Sec. \ref{Sec3b}).

\section{Conclusions}\label{Conclusions}

Information about one-neutron halo nuclei cannot be obtained with direct spectroscopic techniques but are   inferred from indirect methods, such as reactions.  Inclusive breakup reactions are of particular interest since they have much higher statistics than exclusive measurements. To reliably extract  structure information, one needs to know precisely the sensitivity of the reaction observables to the projectile description. In this work, we investigate how the ground-state wavefunction,  the presence of subthreshold  excited states and resonances in the core-neutron continuum influence the parallel-momentum distribution of the remaining core after the collision. We also study the influence of these structure features on the relative core-neutron energy distribution after the diffractive breakup of one-neutron halo nuclei. We perform this analysis for the one-neutron knockout of $^{11}\mathrm{Be}$ on $^{12}\mathrm{C}$ at 68~MeV/nucleon.

By using a Halo-EFT description of $^{11}\mathrm{Be}$~\cite{HJP17,CPH18}, we generate ground-state wavefunctions with very different inner parts but similar large-distance behavior. We show that the parallel-momentum distributions of both the diffractive breakup and stripping, are sensitive  only to  the asymptotics of the ground-state wavefunction. {This confirms the conclusions of Refs.~\cite{H96,BH02} for  knockout and of Ref.~\cite{CN07} for diffractive breakup:} the inclusive breakup observables cannot be used to probe the ground-state wavefunction below 4~fm. In particular, the norm of the overlap wavefunction, i.e., the spectroscopic factor, cannot be determined {reliably} from such observables.
  However,  information about the tail of the wavefunction, viz.\ the ANC, can  be safely extracted.

The presence of an  excited subthreshold state, such as the $1/2^-$ excited state in $^{11}$Be,  reduces the {breakup cross section.
We have demonstrated that at the adiabatic approximation this reduction in the breakup amplitude is  transferred to the inelastic channel, viz.\ to the excitation of the projectile towards that subthreshold state (see Appendix~\ref{App1}). }

We have also shown that the presence of a resonance in the continuum has a negligible impact on the parallel-momentum distribution for inclusive breakup reactions.
Therefore, in the theoretical analyses of these distributions, an accurate description of the core-neutron continuum is not needed.
This strongly reduces the uncertainty related to the projectile model in the study of such reactions.
These inclusive observables are ideal to extract structure information pertaining to the asymptotics of the ground-state wavefunction, such as the ANC~\cite{TetalTAM02}.

A direct application of this work is to reanalyze existing experimental data on $^{11}$Be and $^{15}$C \cite{Aetal00,Tetal02PRC,Fang04} and see if the ANC that can be inferred from these data is in agreement with the \textit{ab initio} calculations of Calci \etal~\cite{Cetal16}.
Hopefully, this would confirm similar analyses performed {recently} for diffractive breakup \cite{CPH18,MC19} and transfer \cite{YC18,Metal19}.
In the future, we plan to extend this idea to two-neutron halo nuclei, using the eikonal framework for three-body projectiles~\cite{CFH17,Setal18}.

\appendix
{\section{Relation between the breakup and inelastic cross sections}\label{App1}}
At the usual eikonal model, i.e., relying on the adiabatic approximation, the energy-distribution of the diffractive breakup reads~\cite{CBS08,GBC06}
\begin{eqnarray}
\derivative{\sigma_{\rm bu}}{ E }&=& \frac{ 4\mu_{cn}}{ \hbar^2 k} \frac{1}{2J_0+1} \sum_{M_0} \sum_{lJM}\int b\,db\left|S_{klJM}^{(M_0)}(b)\right|^2,\label{eqA1}
\end{eqnarray}
where $\ve{b}$ is the transverse coordinate of $\ve{R}$ (see Fig.~\ref{Fig3BodyCoordinates}). The breakup amplitude $ S_{klJM}^{(M_0)}$ is defined from the radial part of the  three-body wavefunction  $\psi_{lJM}^{(M_0)}$ in the $lJM$  wave at $Z\to\infty$   and the radial continuum wavefunctions $u_{lJ}(E)$~\cite{CBS08,GBC06}
 \begin{equation}
 S_{klJM}^{(M_0)}(b)=e^{i(\sigma_l+\delta_{lJ}-l\pi/2)} \int_0^{\infty} dr \, u_{lJ}(E,r) \psi^{(M_0)}_{lJM} (b,r), \label{eqA2}
 \end{equation} 
 where $\sigma_l$ is the Coulomb phase shift.
The contribution to the diffractive-breakup cross section of each partial wave $\sigma_{\rm  bu}^{lJM}$ is simply obtained by integrating the corresponding contribution to the energy distribution~\eqref{eqA1}
\begin{equation}
\sigma_{\rm bu}^{lJM}=\frac{ 4\mu_{cn}}{ \hbar^2 k} \frac{1}{2J_0+1} \sum_{M_0}\int dE\int b\,db\left|S_{klJM}^{(M_0)}(b)\right|^2.
\end{equation}
From the definition of the breakup amplitude~\eqref{eqA2} and the closure relation~\eqref{eq11}, we can write the breakup contribution of a partial wave which does not include the initial ground state, i.e., $lJM\neq l_0J_0M_0$, as
\begin{eqnarray}
\sigma_{\rm bu}^{lJM}&=& \frac{1}{2J_0+1} \sum_{M_0}\int d\ve{b} \int dr |\psi_{lJM}^{(M_0)}(b,r)|^2\nonumber \\
&&-\sum_n \sigma_{\rm inel}^{nlJM},\label{eqA4}
\end{eqnarray}
where 
\begin{equation}
\sigma_{\rm inel}^{nlJM}=\frac{1}{2J_0+1} \sum_{M_0} \int d\ve{b} \left|\int dr\,u_{nlJ}(r) \psi_{lJM}^{(M_0)}(b,r)\right|^2\label{eqA5}
\end{equation} is  the  contribution of the bound state $nlJM$ to the inelastic scattering cross sections. The first term of \eqref{eqA4} does not depend on the description of the continuum, nor on the presence of excited states, the sum 
\begin{equation}
\sigma_{\rm sum}^{lJM}=\sigma_{\rm bu}^{lJM}+\sum_n\sigma_{\rm  inel}^{nlJM} \label{eqA6}
\end{equation} 
is therefore sensitive  only to the ground-state wavefunction and  the optical potentials. This relation explains the transfer within each partial-wave of the flux from the breakup to the inelastic-scattering channel  when an additional bound state is included (see Sec.~\ref{Sec3c}).  It also predicts the complete independence of the integrated cross sections from the description of the continuum, observed in Sec.~\ref{Sec3d}. {Note that this relation is  valid only if the adiabatic approximation holds.}

\begin{acknowledgements}
	C.~Hebborn acknowledges the support of the Fund for Research Training in Industry and Agriculture (FRIA), Belgium.   This project has received funding from the European Union’s Horizon 2020 research	and innovation program under grant agreement 	No 654002, the Deutsche Forschungsgemeinschaft within the Collaborative	Research Centers 1245 and 1044, and the PRISMA+ (Precision Physics, Fundamental Interactions and Structure of Matter) Cluster of Excellence. P. C. acknowledges the support of the State of Rhineland-Palatinate.
\end{acknowledgements}

\bibliographystyle{apsrev}

\end{document}